\documentclass[12pt]{article}
\usepackage{amssymb,amsmath,cite,geometry,graphicx,url}
\catcode`\@=11

\@addtoreset{equation}{section}
\def\theequation{\arabic{section}.\arabic{equation}}
     
\catcode`\@=11
\def\thesection{\arabic{section}.}

\def\appendix{\setcounter{section}{0}
        \def\thesection{Appendix.}
        \def\theequation{\Alph{section}.\arabic{equation}}}
\def\section{\@startsection{section}{1}{\z@}{3.5ex plus 1ex minus
   .2ex}{2.3ex plus .2ex}{\large\bf}}
     
\long\def\@makefntext#1{\parindent 0cm\noindent
\hbox to 1em{\hss$^{\@thefnmark}$}#1}

\newcommand{\captionfonts}{\small}
\makeatletter  % Allow the use of @ in command names
\long\def\@makecaption#1#2{%
  \vskip\abovecaptionskip
  \sbox\@tempboxa{{\captionfonts #1: #2}}%
  \ifdim \wd\@tempboxa >\hsize
    {\captionfonts #1: #2\par}
  \else
    \hbox to\hsize{\hfil\box\@tempboxa\hfil}%
  \fi
  \vskip\belowcaptionskip}
\makeatother   % Cancel the effect of \makeatletter 

\begin{document}
\begin{titlepage}
\vspace{.5in}
\begin{flushright}
September 2017\\  %date
\end{flushright}
\vspace{.3in}
 
\begin{center}
{\Large\bf
 Dimensional reduction in manifold-like causal sets}\\  %title
\vspace{.4in}
{J.~A{\sc bajian}\footnote{\it email: jabajian@trinity.edu}\\
       {\small\it Department of Physics}\\
       {\small\it Trinity University}\\
       {\small\it San Antonio, TX 78212}\\{\small\it USA}}\\[2ex]
{S.~C{\sc arlip}\footnote{\it email: carlip@physics.ucdavis.edu}\\
       {\small\it Department of Physics}\\
       {\small\it University of California}\\
       {\small\it Davis, CA 95616}\\{\small\it USA}}
\end{center}

\vspace{.3in}
\begin{center}
{\large\bf Abstract}
\end{center}
\vspace*{.1ex}
\begin{center}
\begin{minipage}{4.9in}
{\small
We investigate the behavior of small subsets of causal sets that approximate Minkowski
space in three, four, and five dimensions, and show that their effective dimension
decreases smoothly at small distances.   The details of the short
distance behavior depend on a choice of dimensional estimator, but for a reasonable 
version of the Myrheim-Meyer dimension, the minimum dimension is $d \approx 2$,
reproducing results that have been seen in other approaches to quantum gravity.  
}
\end{minipage}
\end{center}
\end{titlepage}
\addtocounter{footnote}{-2}

\section{Introduction}

Despite their enormous differences, many competing approaches to quantum gravity 
share a common feature, a prediction that the effective dimension of spacetime 
decreases near the Planck scale, typically to $d\approx 2$ \cite{Carlipx}.  This 
phenomenon was first noted in high temperature string theory \cite{Atick}, but has 
subsequently  been seen in causal dynamical triangulations \cite{Ambjorn}, asymptotic 
safety \cite{Lauscher,Percacci}, loop quantum gravity \cite{Modesto}, the short distance 
Wheeler-DeWitt equation \cite{Carlipa}, minimum length scenarios \cite{PadChak}, 
and a variety of other approaches; see \cite{Carlipy} for a recent review.

Based in part on the properties of very small causal sets, it was suggested in \cite{Carlipc}
that causal set theory might exhibit similar phenomenon of dimensional reduction.  
Added evidence has come from the short-distance behavior of the causal set 
d'Alembertian \cite{Belenchia} and perhaps from the appearance of ``asymptotic silence'' 
at short distances \cite{Eichhornb}.  But while there are some hints from earlier 
investigations of small subsets of manifold-like causal sets \cite{Reid}, the question 
has not yet been systematically investigated.\footnote{Amusingly, related results 
have been seen in an entirely different context by Clough and Evans \cite{Clough}, 
who use causal set theory to analyze citation networks.}

In this paper, we fill this gap.  Starting with a large sample of Minkowski-like causal sets in 
dimension $d=3$, $4$, and $5$, we evaluate the dimensions of successively smaller
causal diamonds.  As a dimensional estimator, we use the Myrheim-Meyer dimension $d_M$
\cite{Myrheim,Meyer}, essentially a ``box-counting'' dimension adopted to Lorentzian
signature.  There is an ambiguity in the definition of $d_M$, depending on how one 
treats disconnected points.  We find that the dimension of subsets decreases smoothly 
but rapidly at small volume, either to $d_M \approx 2$ for one definition 
or to $d_M \approx 0$ for another.  The former behavior mimics that seen in other 
approaches to quantum gravity.   

\section{Causal sets and Myrheim-Meyer dimension}

A causal set \cite{Bombelli} is the natural Lorentzian version of a discrete spacetime,
a discrete set of events with prescribed causal relations.  Such a set can be described 
by a partial order $\prec$, where $x\prec y$ means ``$x$ is to the past of $y$,'' 
obeying the conditions\\[-4ex]
\begin{enumerate}\addtolength{\itemsep}{-1.5ex}
\item transitivity: $x\prec y$ and $y\prec z \Rightarrow x\prec z$;
\item acyclicity: $x \prec y$ and $y \prec x \Rightarrow x=y$;
\item local finiteness: given any two elements $x$ and $y$, the number of elements 
   $z$ lying between them (i.e.,  $x\prec z\prec y$) is finite.
\end{enumerate}
\vspace*{-1ex}
The causal relations $\prec$ may be viewed as determining ``most'' of the metric.  
In general, the causal structure of a globally hyperbolic manifold determines the metric 
up to a conformal factor \cite{Malamet}; for a causal set, the missing conformal factor 
is simply the number of points in a region.

Most causal sets are not at all manifold-like, and it is an open question whether one 
can find a dynamical principle that limits sets to those that look like nice spacetimes.  
The converse process, however---finding a causal set that approximates 
a given manifold $M$---is straightforward.   Starting with a finite-volume region of a 
globally hyperbolic manifold $M$ with metric $g$, we select  a ``sprinkling'' of points 
by a Poisson process such that the probability of finding $m$ points in any region of 
volume $V$ is 
\begin{align}
P_V(m) = \frac{(\rho V)^m}{m!}e^{-\rho V}
\label{a1}
\end{align}
for a discreteness scale $\rho^{-1}$.  We assign to these points the causal relations 
determined by the metric $g$, and then ``forget'' the original manifold, keeping only 
a set of points and their relations.   At scales larger than $\rho^{-1}$, the resulting 
causal set is expected to be a good approximation of $M$.  In particular, if $M$ is 
Minkowski space, such a causal set preserves statistical Lorentz invariance \cite{LIV}, 
a highly nontrivial result.

In this paper, we will limit ourselves to causal sets obtained from such sprinklings in
Minkowski space.  This is implicitly a dynamical claim: we are assuming that whatever
dynamics underlies causal set theory, it will pick out manifold-like sets.  On large
scales, the quantity we use to measure the dimension requires corrections to 
account for curvature \cite{Roy}, but as long as the curvature scale is much larger 
than the Planck scale, our small-distance results should hold for any manifold-like 
causal set.

As in other approaches to quantum gravity, it is not immediately obvious how to
define the dimension of a causal set \cite{Carlipy}.  Causal sets are inherently 
Lorentzian, and we should presumably look for a ``dimensional estimator'' that
takes this into account.  The standard choice is the Myrheim-Meyer dimension,
which is based on counting causally related points.

Start with a causal set obtained from a Poisson sprinkling on $d$-dimensional 
Minkowski space.  Select an Alexandrov interval, or ``causal diamond,'' $\mathcal{A}$%
---that is, a set consisting of the intersection of the future of a point $p$ and the past 
of another point $q$.  Define $\langle C_1\rangle$ to be the average number of 
points in $\mathcal{A}$, and $\langle C_2\rangle$ to be the average number of causal 
relations, that is, pairs $x,y$ such that $x\prec y$.  The quantities $\langle C_1\rangle$  
and $\langle C_2\rangle$ depend on the volume and the discreteness scale, but a 
suitable ratio depends only on the dimension:
\begin{align}
\frac{\langle C_2\rangle\,}{\langle C_1\rangle^2} =
   \frac{\Gamma(d+1)\Gamma(\frac{d}{2})}{4\Gamma(\frac{3d}{2})}
\label{b1}
\end{align}
The right-hand side of (\ref{b1}) is a monotonic function, and can be inverted.
For an arbitrary causal set, the Myrheim-Meyer dimension $d_M$ is then defined to
be the value $d$ for which (\ref{b1}) holds.   

As noted earlier, there is an ambiguity in this definition when a causal set contains an
isolated point, a point with no causal relations with any others.  A single point should
presumably have dimension $d=0$, but the left-hand side of (\ref{b1}) is zero for 
such a point, which would correspond to $d\rightarrow\infty$ on the right-hand
side.  There seem to be two natural ways to treat such isolated points: we can ascribe
a dimension of zero to them, or we can simply neglect them, on the grounds that
a completely causally disconnected point is not really part of spacetime.  For large
causal sets, the choice  makes no appreciable difference to the Myrheim-Meyer 
dimension, but as we shall see, for small enough sets it matters.

\section{Approach}

\begin{figure}
\begin{center}
\includegraphics[height=2.5in]{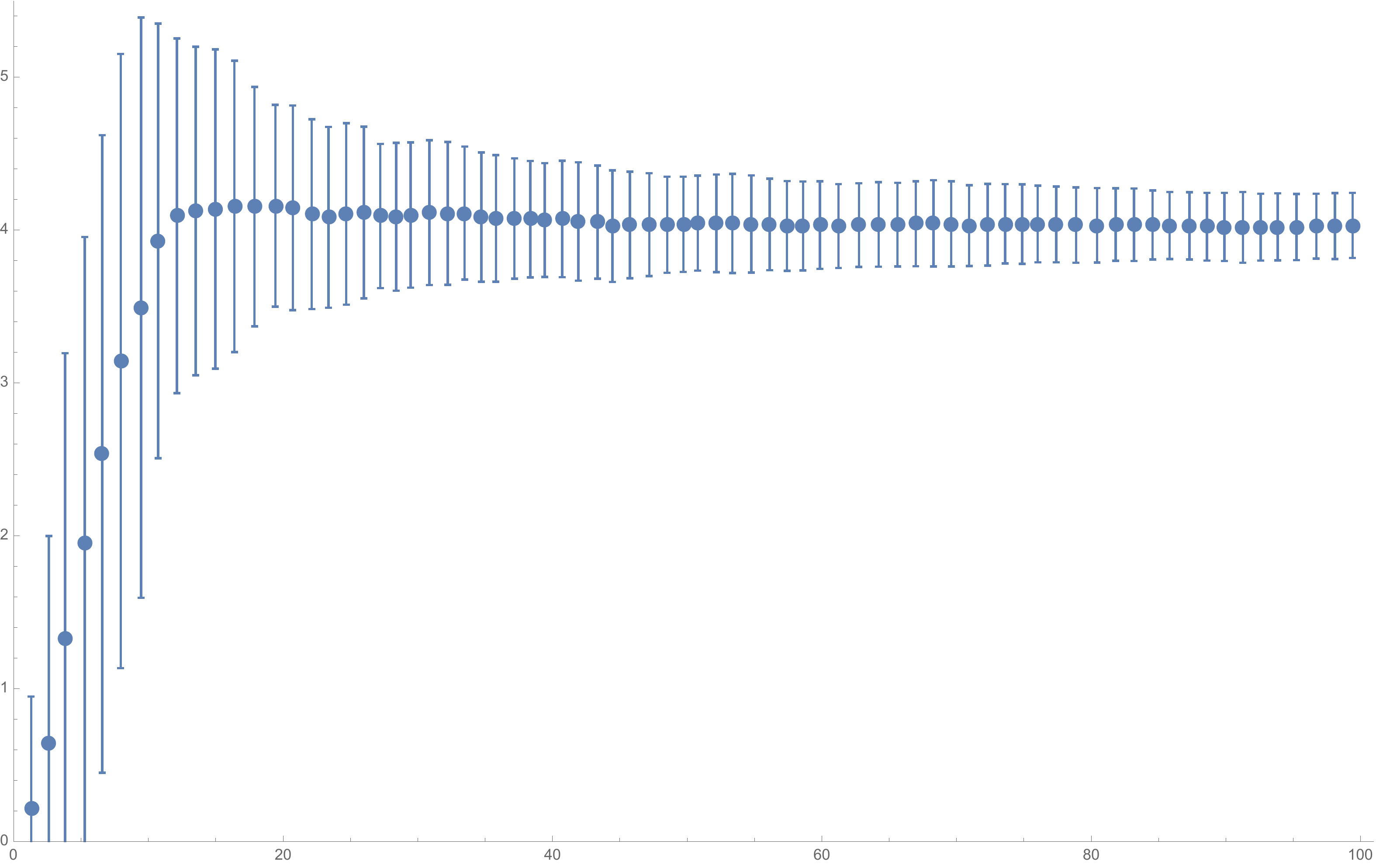}
\caption{Myrheim-Meyer dimension for subsets of a relatively large causal set \label{fig1}}
\end{center}
\end{figure}

To generate the causal sets used in our analysis, we created a Mathematica notebook 
\cite{Math} that allowed us to select random points uniformly from a causal diamond in 
Minkowski space, initially in four dimensions.   We calculated the Myrheim-Meyer dimension 
and verified that it agreed with the dimension of of the background manifold in the  limit 
of dense sprinklings.  As is evident from figure \ref{fig1}, this limit was already reached
with sprinklings of about 20 points, so we used this as a typical size.

To investigate the dependence of dimension on volume, one must choose a way to 
of select causal sets of ``small" volume.  Since the volume of a causal set is 
determined by the number of points in the set, it is tempting to simply average 
over all subsets containing a given number of points.  This can be misleading, though: 
while such sets are ``small" if viewed outside the context of the background  
spacetime, most of them do not come from a small region of the background spacetime, 
but include points spread across a large region of the background manifold.   In
particular, two points with a lightlike separation can be ``adjacent'' in a causal set
even if they are widely separated in spacetime.

As an alternative, for each of our sprinklings we considered successively smaller sub-diamonds 
in the background spacetime.  The points in each sub-diamond constitute a new causal 
set, whose volume and Myrheim-Meyer dimension we computed.  We repeated the 
process for 10,000 sprinklings, and then averaged the dimension at 
each volume.  We initially applied this analysis to four-dimensional Minkowski space,
but subsequently repeated it for $d=3$ and $d=5$.

As noted above, the Myrheim-Meyer dimension is not well-defined for single points or
causal sets with no edges.  While this concern is unimportant for large causal sets, it
must be confronted for the very small sets  we are interested in.  We explored 
two reasonable possibilities: taking the volume of an isolated point to be zero (they are,
after all, single points) or dropping edgeless causal sets from our counting (they
are causally disconnected from the rest of spacetime).

\section{Results}

\begin{figure}
\begin{center}
\includegraphics[width=3.15in]{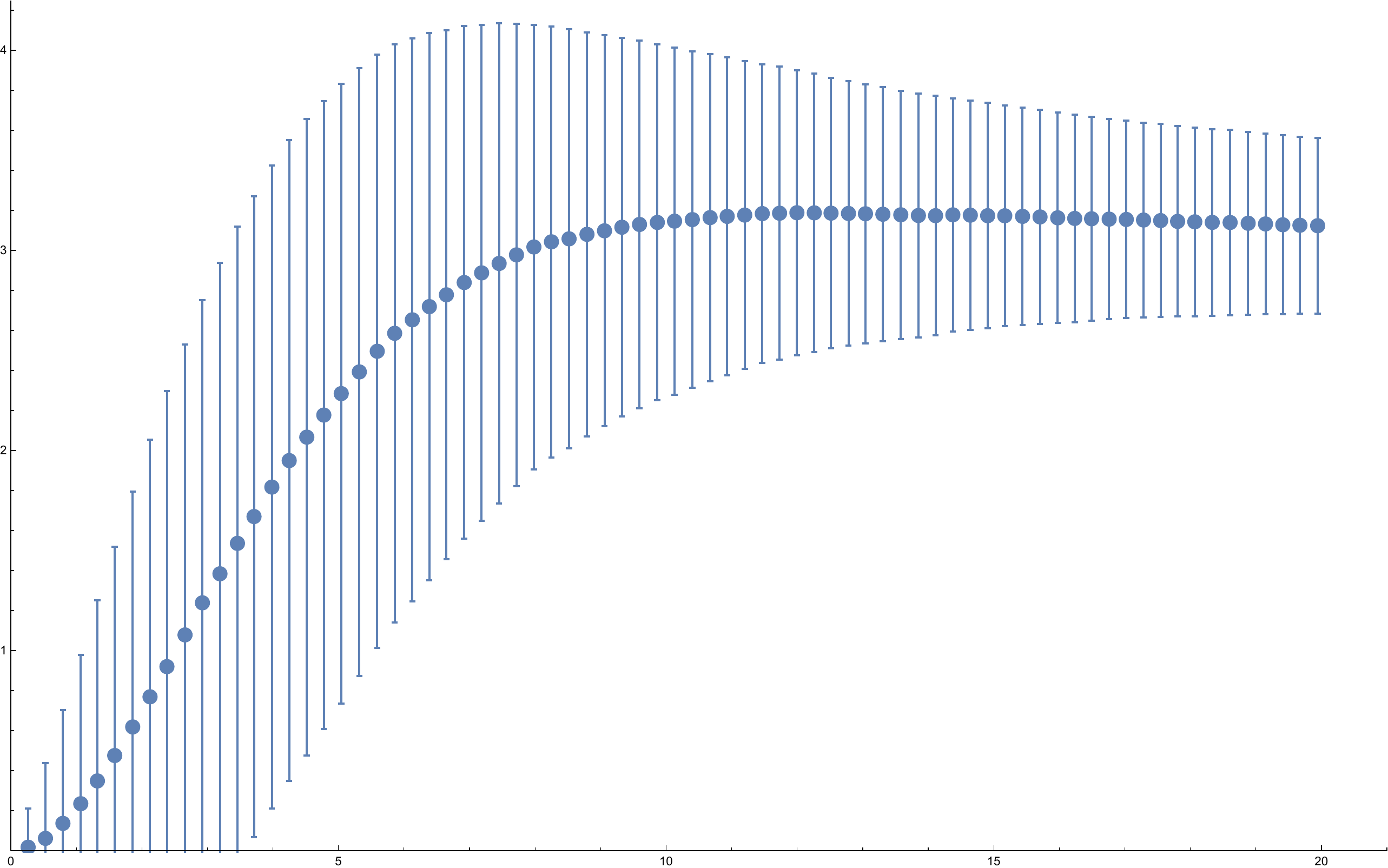}
 \includegraphics[width=3.15in]{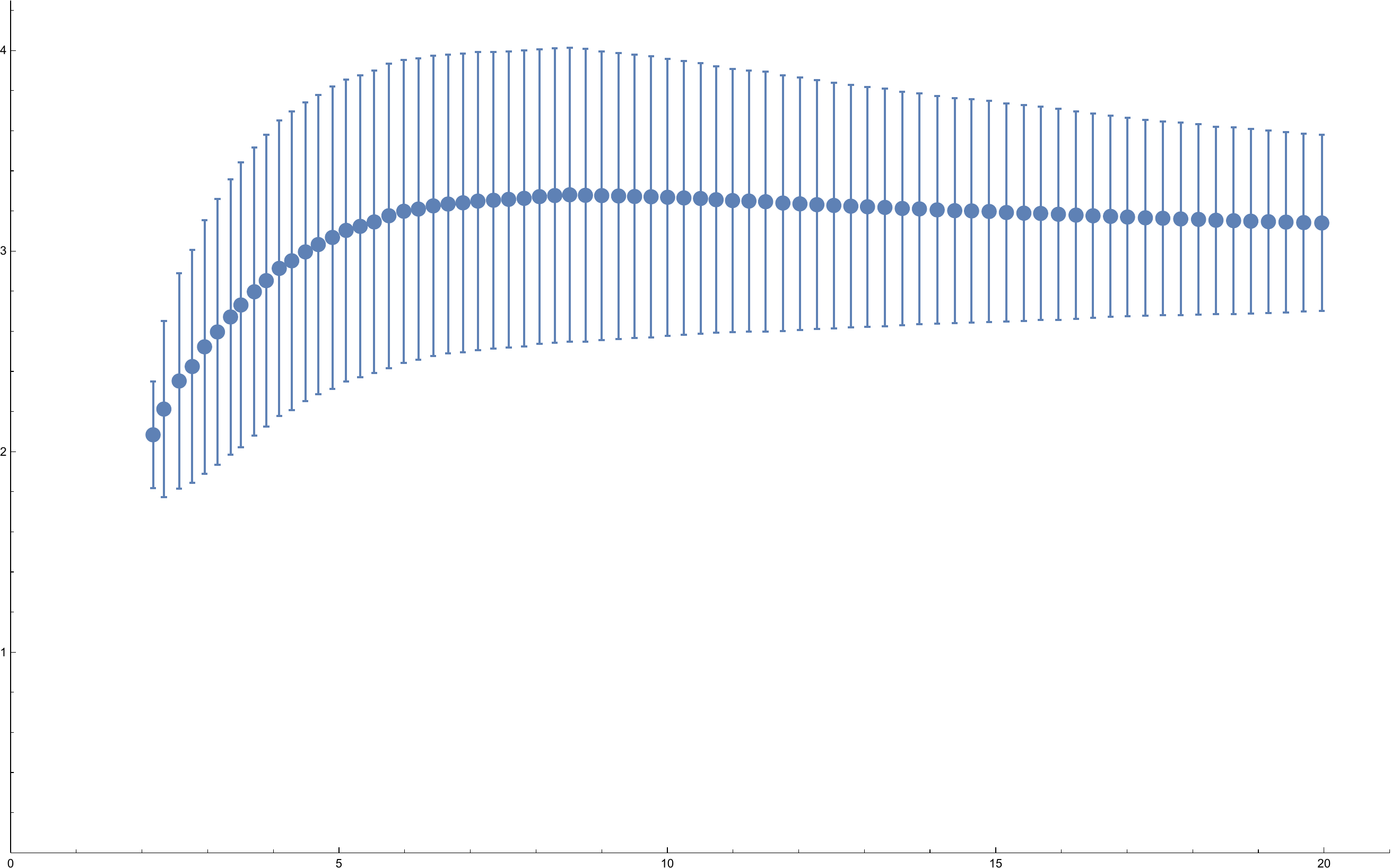}
\caption{Myrheim-Meyer dimension in a three-dimensional background,
with edgeless sets counted as dimension zero (left) or omitted (right) \label{fig2}}
\end{center}
\begin{center}
\includegraphics[width=3.15in]{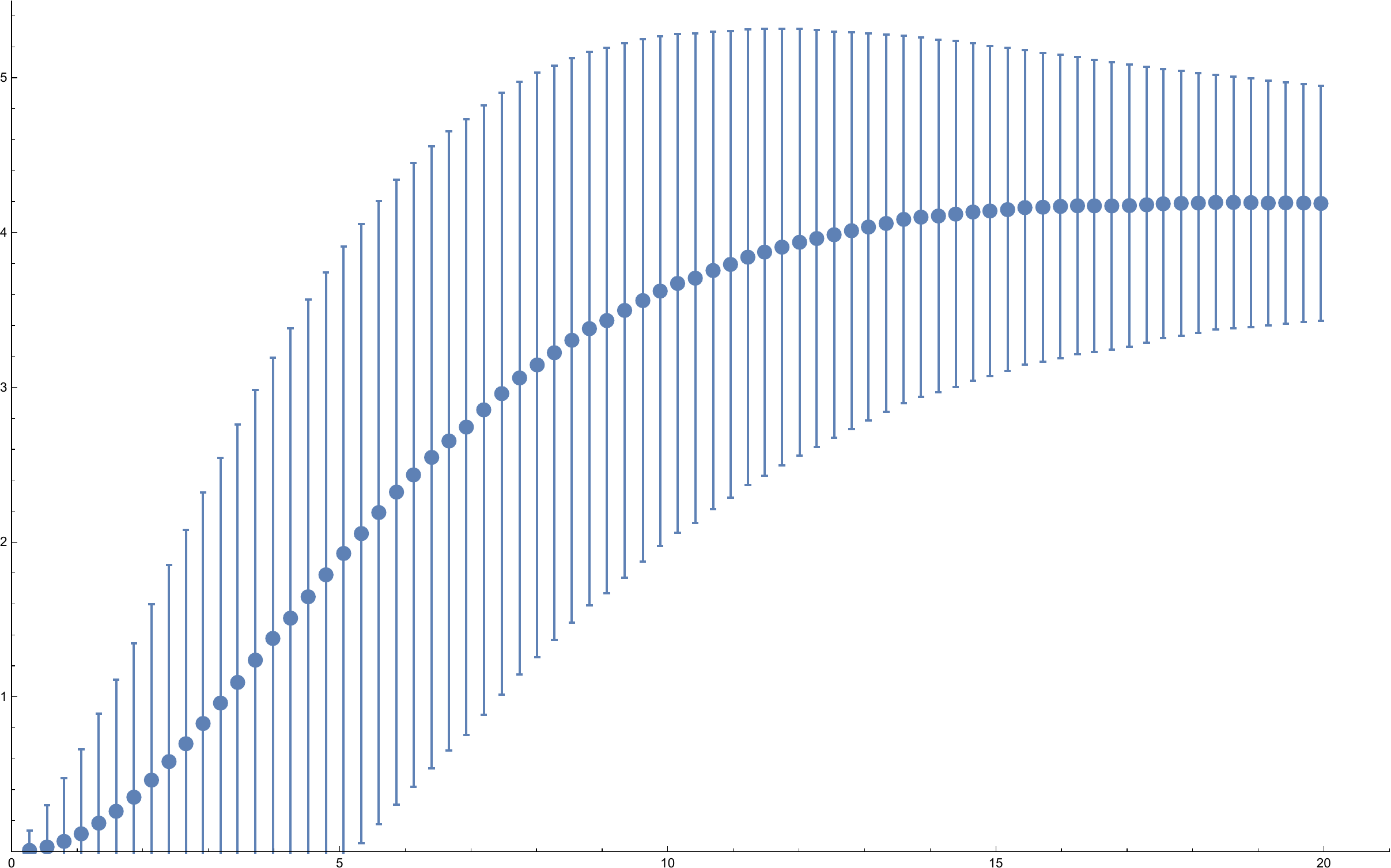}
 \includegraphics[width=3.15in]{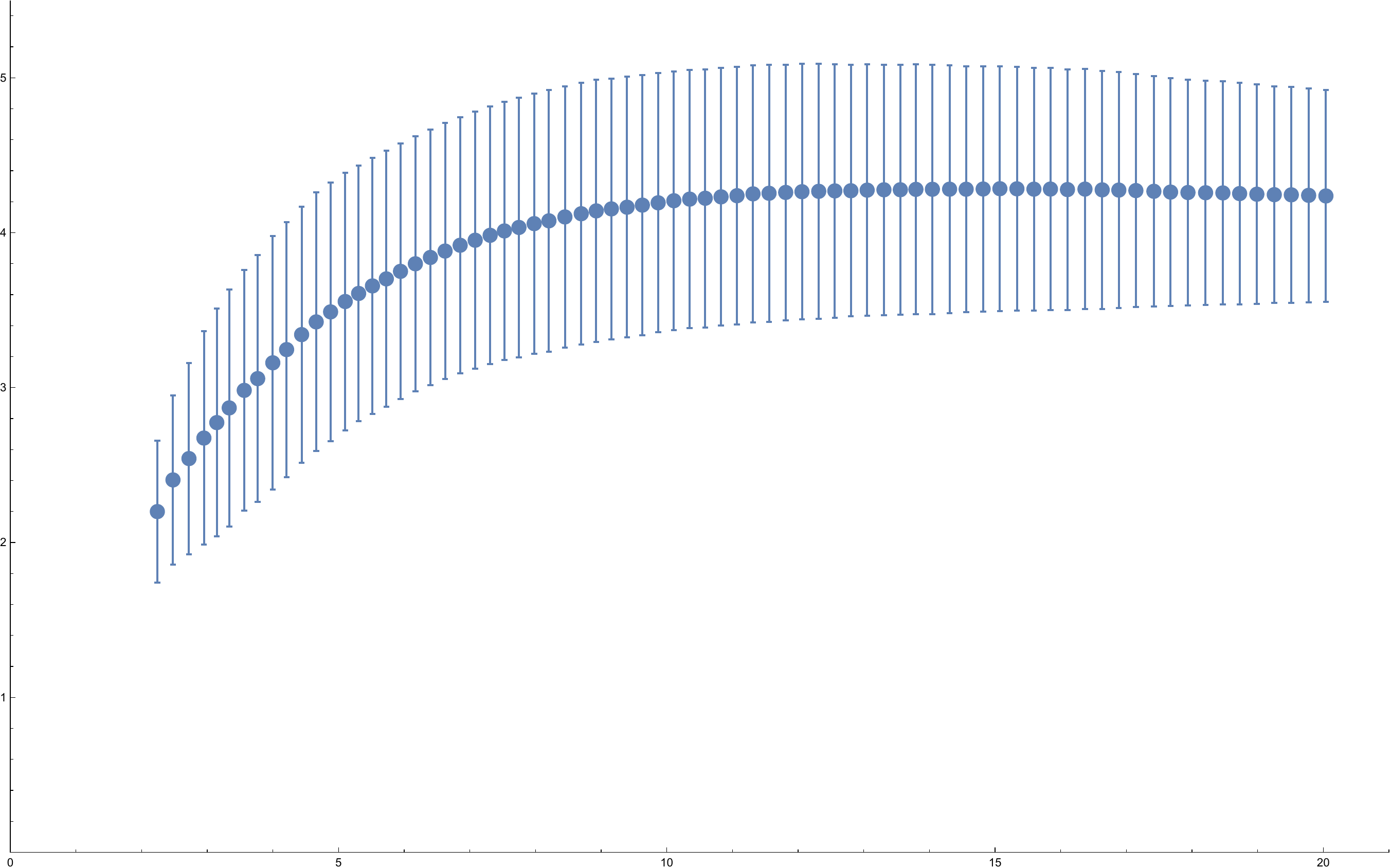}
\caption{Myrheim-Meyer dimension in a four-dimensional background,
with edgeless sets counted as dimension zero (left) or omitted (right) \label{fig3}}
\end{center}
\begin{center}
\includegraphics[width=3.15in]{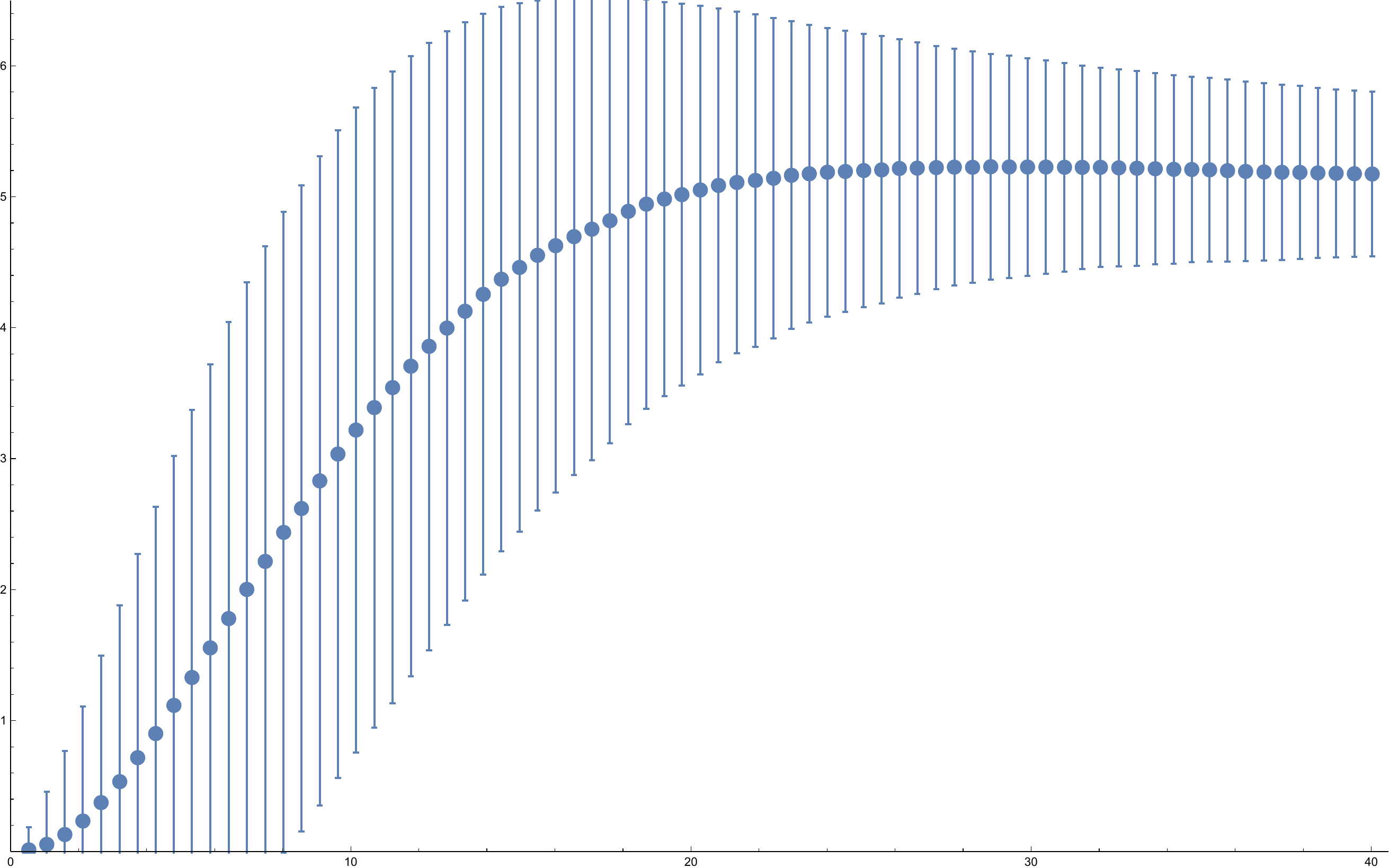}
 \includegraphics[width=3.15in]{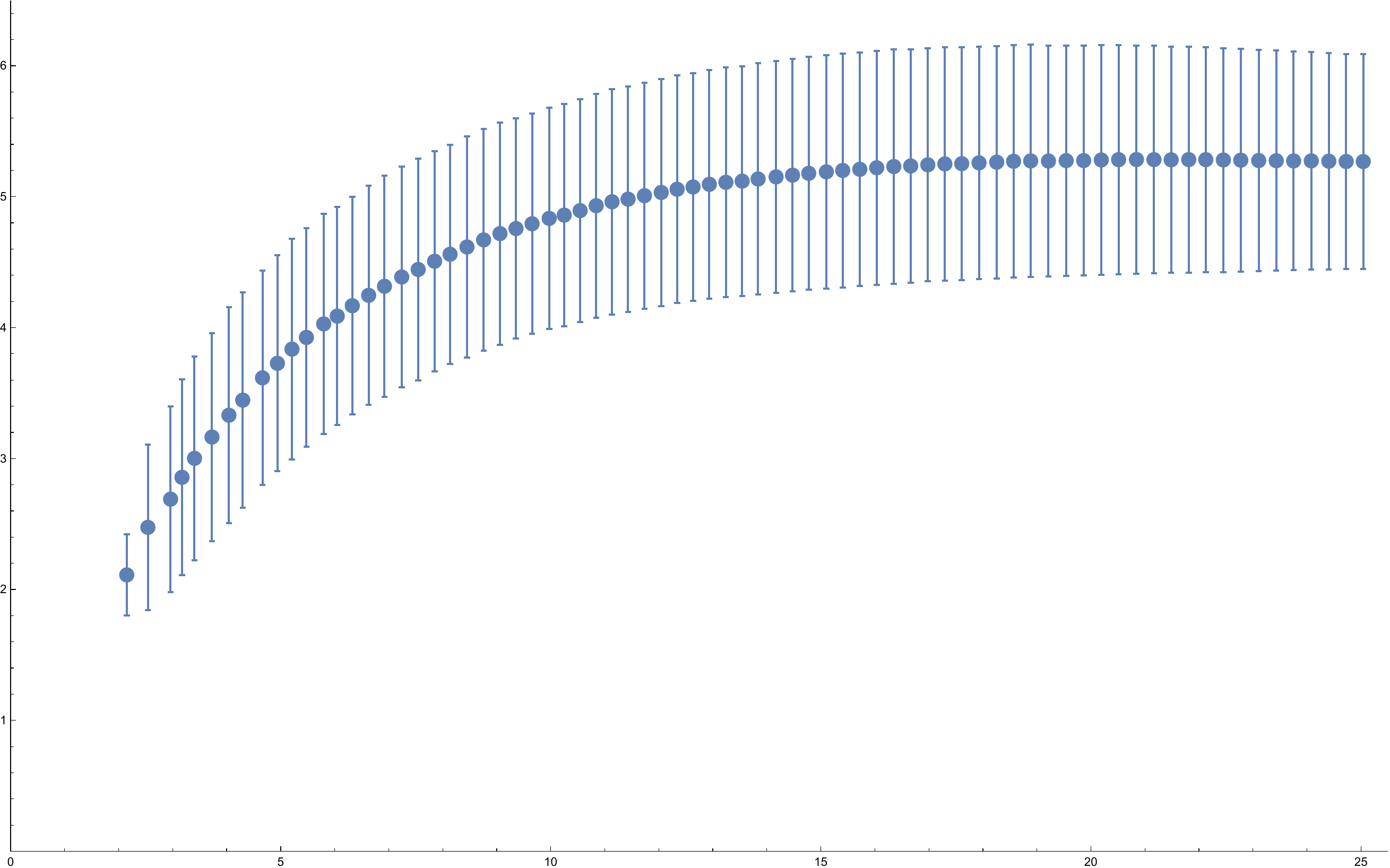}
\caption{Myrheim-Meyer dimension in a five-dimensional background,
with edgeless sets counted as dimension zero (left) or omitted (right) \label{fig4}}
\end{center}
\end{figure}

In each of the background dimensions we studied, we found that dimensional reduction 
does indeed occur as the volume decreases.   As shown in figures \ref{fig2}--\ref{fig4},  
the process appears to be smooth, but has a rather abrupt onset.  The 
transition to lower dimension starts at a volume of approximately $V=8$ points in 
three dimensions, $V=16$ points in four dimensions, and $V=22$ points in five 
dimensions.\footnote{Fractional volumes appear in the graphs because at a given
background volume in Minkowski space, causal sets with varying numbers of points 
may be present.}

At volumes above the transition, the Myrheim-Meyer dimension remains stable and
equal to the dimension of the background Minkowski space.  Below the transition,
the decrease is quite rapid.  For each of the background dimensions we considered, 
the minimum Myrheim-Meyer dimension falls to $d_M\approx 0$ if edgeless causal
sets are taken to have dimension zero, and $d_M\approx 2$ if they are omitted.
We can understand the latter result by noting that the smallest causal set with an
edge---two points with one relation---has a Myrheim-Meyer dimension of two. 

Figures \ref{fig2}--\ref{fig4} show $1\, \sigma$ error bars.  We believe these are not 
a result of poor statistics, but are rather a consequence of our definition of
volume.  A causal diamond of a given volume in a background Minkowski space can
contain many different causal sets, which will not all have identical Myrheim-Meyer
dimensions.  This leads to a genuine statistical fluctuation in dimension, especially
at small volumes.

The end point $d_M\approx 2$ is reminiscent of the behavior seen in other 
investigations of quantum gravity \cite{Carlipy}.  More precisely, when edgeless causal 
sets are discarded, we find a minimum dimension of $d_M = 2.08 \pm .26$ in three 
background dimensions, $d_M = 2.13 \pm .39$ in four background dimensions, and 
$d_M = 2.19 \pm .40$ in five background  dimensions.  It would be interesting to 
understand the fluctuations better, especially since a few other approaches to 
quantum gravity suggest a minimum dimension of $3/2$ \cite{Coumbe,Laiho}
or $5/2$ \cite{Husain,Ronco}.

We would also like to understand what determines the scale at which dimensional
reduction sets in.  For three and four background dimensions, the transition seems to 
occur at a characteristic length of about twice the sprinkling length---that is, 
$V\sim 2^d$ points---but this pattern appears to break down for background
 dimension five.  We also plan to investigate the behavior of another standard
dimensional estimator, midpoint scaling dimension \cite{Reid}.

Ideally, we would like to do more.  The results we have presented here have the 
awkward feature of relying on the background Minkowski space to define the
small regions whose dimension we measure.  This was necessary to avoid picking 
out causal subsets that were ``small'' in the sense of having few points, but
``large'' in the sense of occupying a highly extended region.  Recently, some
progress has been made in defining ``local'' regions entirely in the context of
causal sets, without reference to any background \cite{Glaser}.  It might be
possible to use this work to investigate dimensional reduction more intrinsically.

\newpage

\vspace{1.5ex}
\begin{flushleft}
\large\bf Acknowledgments
\end{flushleft}

S.~C.\ was supported in part by Department of Energy grant DE-FG02-91ER40674.  J.~A.\ was
supported in part by National Science Foundation grant PHY-1560482.


\begin{thebibliography}{99}
\bibitem{Carlipx} S.\ Carlip,  Int.\ J.\ Mod.\ Phys.\ D25 (2016) 1643003, arXiv:1605.05694.
\bibitem{Atick} J.~J.\ Atick and E.\ Witten,  Nucl.\ Phys. B310 (1988) 291.
\bibitem{Ambjorn} J.~Ambj{\o}rn, J.~Jurkiewicz, and R.~Loll,  Phys.\ Rev.\ Lett.\    95 (2005) 
    171301, arXiv:hep-th/0505113. 
\bibitem{Lauscher} O.\ Lauscher and M.\ Reuter, Phys.\ Rev.\ D65 (2002) 025013, arXiv:hep-th/0108040.
\bibitem{Percacci} R.~Percacci and D.~Perini, Class.\ Quant.\ Grav.\ 21 (2004) 5035, arXiv:hep-th/0401071.
\bibitem{Modesto} L.\ Modesto, Class.\ Quant.\ Grav.\ 26 (2009) 242002, arXiv:0812.2214.
\bibitem{Carlipa} S.\ Carlip, in \emph{Proc.\ of the 25th Max Born Symposium: The Planck 
     Scale}, AIP Conf.\ Proc.\ 1196 (2009) 72,  arXiv:1009.1136.
\bibitem{PadChak} T.\ Padmanabhan, S.\ Chakraborty, and D.\ Kothawala, Gen.\ Rel.\ Grav.\ 48 (2016)  
   55, arXiv:1507.05669.
\bibitem{Carlipy} S.\ Carlip, Class.\ Quant.\ Grav.\ 34 (2017) 193001arXiv:1705.05417.
\bibitem{Carlipc} S.\ Carlip, Class.\ Quant.\ Grav.\ 32 (2015) 232001, arXiv:1506.08775.
\bibitem{Belenchia} A.\ Belenchia, D.~M.~T.\ Benincasa, A.\ Marciano, and L.\ Modesto, Phys.\ Rev.\
    D93 (2016) 044017, arXiv:1507.00330.
\bibitem{Eichhornb} A.\ Eichhorn, S.\ Mizera, and S.\ Surya, arXiv:1703.08454.
\bibitem{Reid} D.~D.\ Reid, Phys.\ Rev.\ D67 (2003) 024034, arXiv:gr-qc/0207103.
\bibitem{Clough} J.~R.\ Clough and T.~S.\ Evans, Physica A 448 (2016) 235, arXiv:1408.1274.
\bibitem{Myrheim} J.~Myrheim, 1978 CERN preprint TH-2538.
\bibitem{Meyer} D.~A.~Meyer, Ph.D.\ thesis, MIT (1989), \url{http://hdl.handle.net/1721.1/14328}.
\bibitem{Bombelli} L.~Bombelli, J.~Lee, D.~Meyer, and R.~Sorkin,    Phys.\ Rev.\ Lett.\ 59 (1987) 521.
\bibitem{Malamet} D.~Malamet, J.\ Math.\ Phys.\ 18 (1977) 1399.
\bibitem{LIV} L.~Bombelli, J.~Henson, and R.~D.~Sorkin, Mod.\ Phys.\ Lett.\
   A24 (2009) 2579, arXiv:gr-qc/0605006.
\bibitem{Roy} M.~Roy, D.~Sinha, and S.~Surya, Phys.\ Rev D87 (2013)
   044046 arXiv:1212.0631.
\bibitem{Math} Wolfram Research, Inc., Mathematica, Version 11.1, Champaign, IL (2017).
\bibitem{Coumbe} D.\ Coumbe and J.~Jurkiewicz, JHEP 1503 (2015) 151, arXiv:1411.7712.
\bibitem{Laiho} 	J.\ Laiho, S.\ Bassler, D.\ Coumbe, D.\ Du, and J.~T.\ Neelakanta, arXiv:1604.02745.
\bibitem{Husain} V.\ Husain, S.~S.\ Seahra, and E.~J.\ Webster, Phys.\ Rev.\ D88 (2013) 024014, 
   arXiv:1305.2814.
\bibitem{Ronco} M.\ Ronco, Adv.\ High Energy Phys.\ 2016 (2016) 9897051, arXiv:1605.05979.
\bibitem{Glaser} L.\ Glaser and S.\ Surya, Phys.\ Rev.\ D 88 (2013) 124026, arXiv:1309.3403.



\end{thebibliography}
\end{document}